\documentclass[12pt,letterpaper,english,preprint, aps, prd, 11pt]{revtex4-1}
\usepackage[T1]{fontenc}
\usepackage[latin9]{inputenc}
\usepackage[dvipdf]{graphicx}

\makeatletter

\providecommand{\tabularnewline}{\\}

\makeatother

\usepackage{babel}
\begin{document}

\title{{\normalsize{}Probing the Anomalous FCNC $tq\gamma$ Couplings at
Large Hadron electron Collider}}

\author{{\normalsize{}I. Turk Cakir}}
\email{ilkay.turk.cakir@cern.ch}

\affiliation{Giresun University, Department of Energy Systems Engineering, 28200,
Giresun, Turkey}

\affiliation{CERN, European Organization for Nuclear Research, Geneva, Switzerland}

\author{{\normalsize{}A. Yilmaz}}
\email{aliyilmaz@giresun.edu.tr}

\affiliation{Giresun University, Department of Electric and Electronics Engineering,
28200, Giresun, Turkey}

\author{{\normalsize{}H. Denizli}}
\email{denizli_h@ibu.edu.tr}

\affiliation{Abant Izzet Baysal University, Department of Physics, 14280, Bolu,
Turkey}

\author{{\normalsize{}A. Senol}}
\email{senol_a@ibu.edu.tr}

\affiliation{Abant Izzet Baysal University, Department of Physics, 14280, Bolu,
Turkey}

\author{{\normalsize{}H. Karadeniz}}
\email{hande.karadeniz@giresun.edu.tr}

\affiliation{Giresun University, Department of Energy Systems Engineering, 28200,
Giresun, Turkey.}

\author{{\normalsize{}O. Cakir}}
\email{ocakir@science.ankara.edu.tr}

\affiliation{Ankara University, Department of Physics, 06100, Tandogan, Ankara,
Turkey}

\date{{\normalsize{}\today}}
\begin{abstract}
{\normalsize{}We investigate the anomalous flavour changing neutral
current (FCNC) interactions of top quark through the process $e^{-}p\to e^{-}W^{\pm}q+X$.
We calculate the signal and background cross sections in electron
proton collisions at Large Hadron electron Collider (LHeC) with a
7 TeV proton beam from the LHC and a new 60 GeV electron beam from
energy recovery linac (ERL). We study the relevant background processes
including one electron and three jets in the final state. The distributions
of the invariant mass of two jets and an additional jet tagged as
$b$-jet are used to account signal and background events after the
analysis cuts. We find upper bounds on anomalous FCNC couplings $\lambda$
of the order of $10^{-2}$ at LHeC for a luminosity projection of
$500$ fb$^{-1}$ together with the fast simulation of detector effects.
As a matter of interest, we analyze the sensitivity to the couplings
$(\lambda_{u},\lambda_{c})$ and find an enhanced sensitivity to $\lambda_{c}$
at the LHeC when compared to the results from the HERA.}{\normalsize \par}
\end{abstract}

\pacs{12.39.Hg Heavy quark effective theory, 13.87.Ce Production, 14.65.Ha
Top quarks.}
\maketitle

\section{{\normalsize{}Introduction}}

The top quark is the most massive of all observed elementary particles.
The top quark interacts with the other quarks of Standard Model (SM)
via gauge and Yukawa couplings. It interacts primarily by the strong
interaction, but can only decay through the weak interaction. The
top quark decays to a $W$ boson and either a bottom quark (most frequently),
a strange quark, or a down quark (rarely). It is also a unique probe
to search for the dynamics of electroweak symmetry breaking. Being
the heaviest quark, effects of new physics on its couplings are expected
to be larger than that for other fermions, hence it is expected that
possible deviations with respect to predictions of the SM might be
found with precise measurements of its couplings. It is known that
the Flavour Changing Neutral Current (FCNC) transitions are absent
at tree level and highly suppressed at loop level due to the Glashow-Illiopoluos-Maiani
(GIM) mechanism \citep{GIM70}. The branching ratios for the top quark
FCNC decays through the process $t\to q\gamma$ are the order of $10^{-14}-10^{-12}$
in the SM \citep{Saavedra04}. However, various scenarios beyond the
standard model (BSM), such as two Higgs doublet model, supersymmetry,
technicolor, predict much larger rates at the order of $10^{-6}-10^{-5}$
\citep{Couture97,Lu03}. Therefore, an observation of the large FCNC
induced couplings would indicate the existence of the BSM.

The experimental limits on the branching ratios of the rare top quark
decays were established by the experiments at Tevatron and LHC. The
CDF experiment set upper bound on branching ratio for top quark FCNC
decay as $BR(t\to q\gamma)<3.2\times10^{-2}$ \citep{CDF98}. Presently,
the most powerful upper limits on top quark FCNC branching ratios
from different channels are $BR(t\to u\gamma)<1.61\times10^{-4}$
and $BR(t\to c\gamma)<1.82\times10^{-3}$ at $95\%$ confidence level
(CL) given by the CMS experiment \citep{CMS15}. 

The direct single top quark production channel through the photon
emission at the LHC have been studied to probe the top quark FCNC
$tq\gamma$ interactions \citep{Jeneret2008,Sun2014}. Search for
flavour changing neutral currents in the single top association with
a photon at the LHC has been studied in Ref. \citep{Guo2016}. In
the same sign top quark production at the LHC the top FCNC couplings
have been studied in Ref. \citep{Reza2015}. The top FCNC couplings
at future circular hadron electron colliders have been studied in
Ref. \citep{Denizli2016}.

The main physics goals of future high energy particle colliders are
to examine the fundamental interactions within the SM and to search
for possible effects of new physics beyond the SM. The electron-proton
colliders are based on a ring type proton accelerator with an intersecting
electron beam accelerator. Currently, a projected future $ep$ collider,
namely Large Hadron electron Collider (LHeC) have been discussed \citep{LHeC,Klein2016,J.L.Abelleira}.
It comprises a 60 GeV electron beam that will collide with the 7 TeV
proton beam, having an integrated luminosity of $L_{int}=100$ fb$^{-1}$
per year and it is planned to collect $1$ ab$^{-1}$ over the years.
Recently, the new physics capability and potential of $ep$ colliders
through di-Higgs boson production have been studied in Ref. \citep{Mukesh}.

When we study anomalous top quark FCNC interactions at the future
colliders, we take also into account bounds from measured low energy
processes where the loops include top quarks. For example, the FCNC
transition processes $b\rightarrow s\gamma$ is considered to be the
valuable probe of top quark anomalous couplings \citep{Yuan2011,Li2011}.
The bounds \citep{Yang2014} on the top quark real FCNC couplings
are lower than the current direct limit but still accessible at the
high luminosity run of LHC.

In this work, we study the physics potential of LHeC collider two
parameters for probing top quark $tq\gamma$ FCNC couplings through
single top production. We focus on two parameters analysis of the
signal and background including Delphes detector simulation. We discuss
the sensitivity and bounds for anomalous top quark couplings ($\lambda_{u},$
$\lambda_{c}$) for different luminosity projections.

\section{{\normalsize{}Calculation Framework}}

\begin{figure}
\includegraphics[scale=0.5]{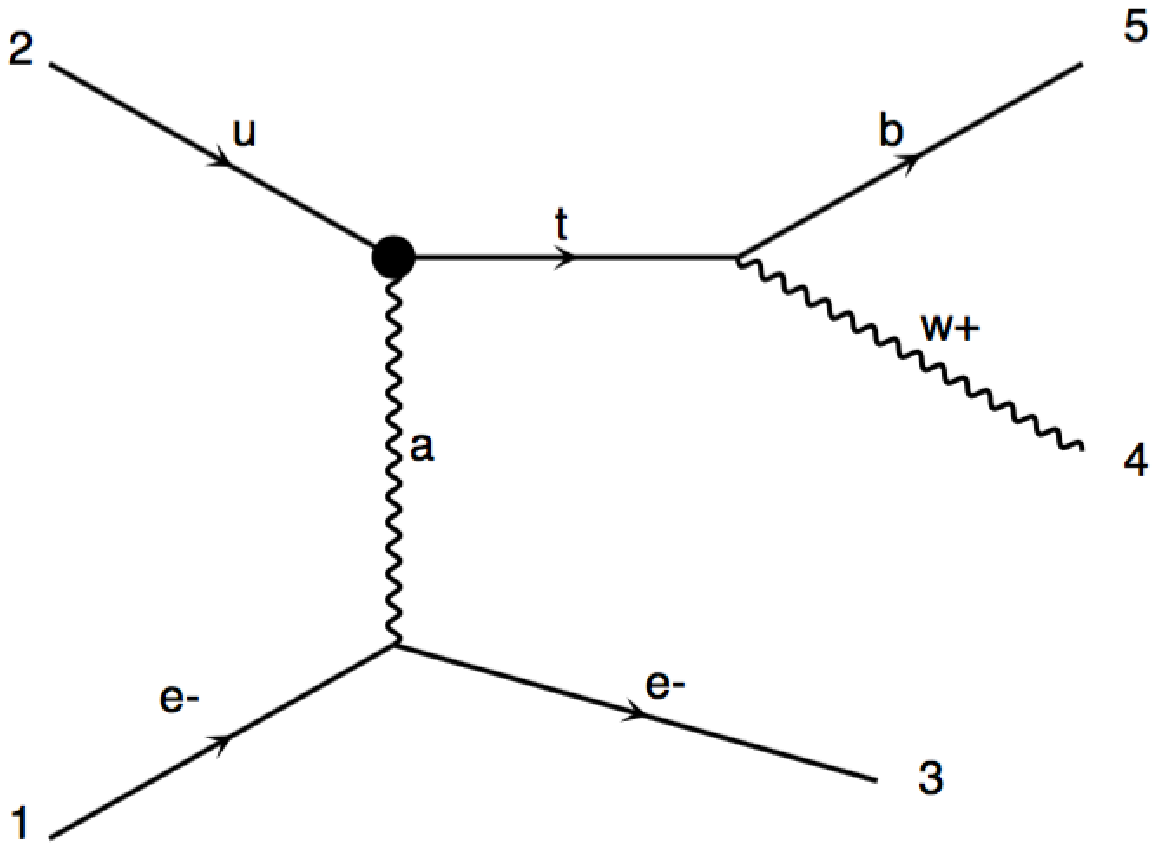}\hspace{0.8cm}\includegraphics[scale=0.5]{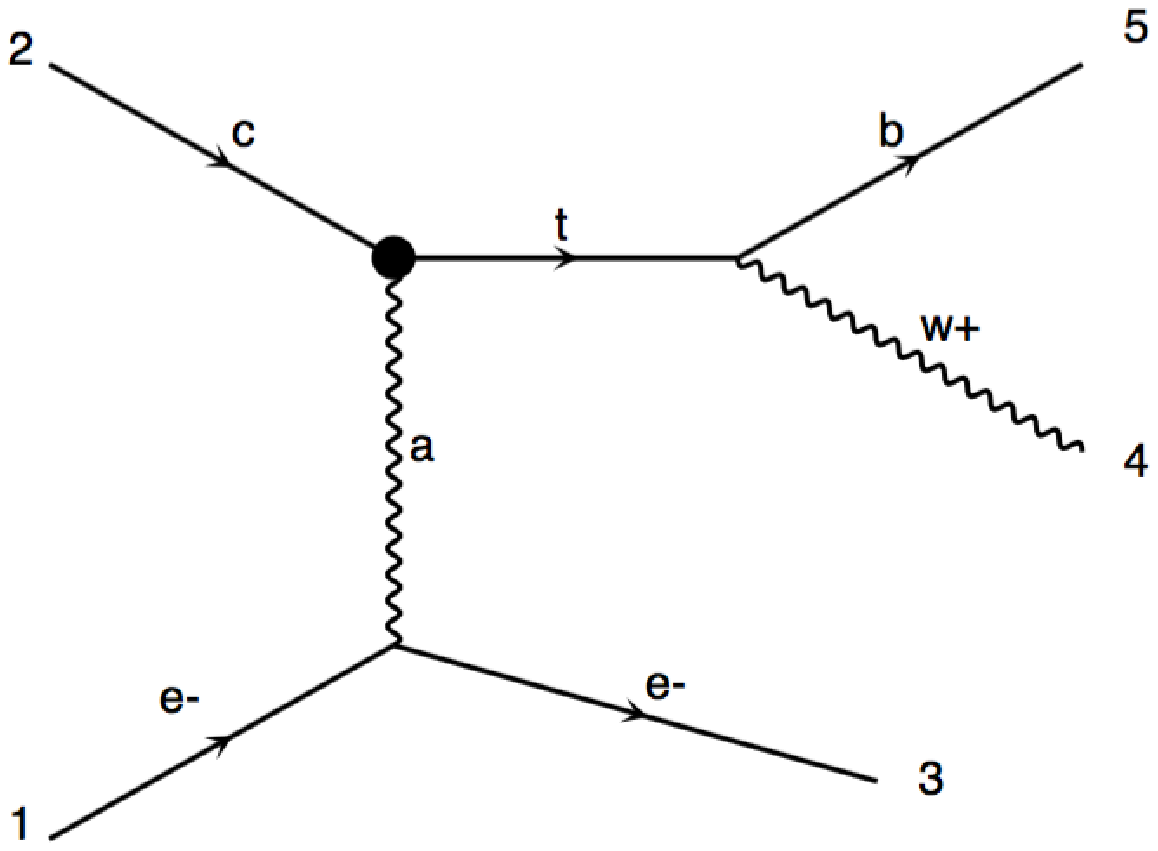}

\caption{Leading order diagrams for single top quark production via flavour
changing neutral currents and its subsequent decay. The similar diagrams
for single anti-top production can be obtained by exchanging $u\leftrightarrow\bar{u}$
and $c\leftrightarrow\bar{c}$ and its corresponding subsequent decay.
\label{fig:fig1}}
\end{figure}

The top quark FCNC interactions would be a good test of new physics
at present and future colliders. The new physics effects can be described
by a set of higher order effective operators in a model independent
framework. The anomalous FCNC interactions of top quark with up-type
quarks ($u$,$c$) and a photon can be described in a model independent
effective Lagrangian 

\begin{equation}
L_{eff}=\frac{g_{e}}{2m_{t}}\bar{t}\sigma^{\mu\nu}(\lambda_{u}^{L}P_{L}+\lambda_{u}^{R}P_{R})uA_{\mu\nu}+\frac{g_{e}}{2m_{t}}\bar{t}\sigma^{\mu\nu}(\lambda_{c}^{L}P_{L}+\lambda_{c}^{R}P_{R})cA_{\mu\nu}+h.c.\label{eq:1}
\end{equation}
 where $g_{e}$ is the electromagnetic coupling constant, $\lambda_{q}^{L(R)}$
are the strength of anomalous FCNC couplings and the values of these
vanish at the lowest order in the SM, $P_{L(R)}$ denotes the left
(right) handed projection operators. The photon field strength tensor
is $A_{\mu\nu}$ and anomalous term is related to $\sigma^{\mu\nu}=\frac{i}{2}[\gamma^{\mu},\gamma^{\nu}]$.
The effective Lagrangian is used to calculate both production cross
section and the decay widths for the $t\to q\gamma$ channels. At
the $ep$ colliders anomalous single top quark production can be achieved
through the scattering of incoming electron beam with the quark current
of the proton via the production channel ($t$-channel) as shown in
diagrams Fig. \ref{fig:fig1}. For the event generation and cross
section calculations, we have used event generator MadGraph5\_aMC@NLO
\citep{Alwall11} with built in parton distribution function NNPDF23
\citep{NNPDF2013}. The effective interactions described by Eq. \ref{eq:1}
have been implemented through FeynRules \citep{Alloul14} and the
model file is used within MadGraph5\_aMC@NLO. This event generator
provides the tools to perform the simulation of both signal and backgrounds
within the same framework. In this study, an allowed range of values
for the FCNC parameters $\lambda_{q}^{L(R)}\leq0.05$ are taken into
account.

\section{{\normalsize{}Cross Section for the Signal }}

The contributions from new interactions of the BSM physics can modify
the production rate and properties of the top quark expected within
the SM. The relative contribution of the left and right currents are
determined by $\lambda_{q}^{L}$ and $\lambda_{q}^{R}$ as given in
Eq. 1, no specific chirality is assumed for the FCNC interaction vertices
($tq\gamma$), here we use a simplified scenario with $\lambda_{q}^{L}=\lambda_{q}^{R}=\lambda_{q}$
throughout the study. We study single top quark (or anti-top quark)
production through the FCNC interactions at the LHeC. The signal cross
sections for the processes $e^{-}p\to e^{-}t+X$ and $e^{-}p\to e^{-}\bar{t}+X$
are $\sigma(t)=196.90$ fb and $\sigma(\bar{t})=40.21$ fb for equal
coupling scenario $\lambda=\lambda_{u}=\lambda_{c}$$=0.05$ at the
center of mass energy $\sqrt{s_{ep}}\simeq1.29$ TeV of the LHeC for
the electron beam energy of $60$ GeV and proton beam energy of $7$
TeV, respectively. In order to enrich the statistics, we take into
account both single top and single anti-top production processes.

\begin{figure}
\includegraphics[clip,width=9.5cm,height=7cm]{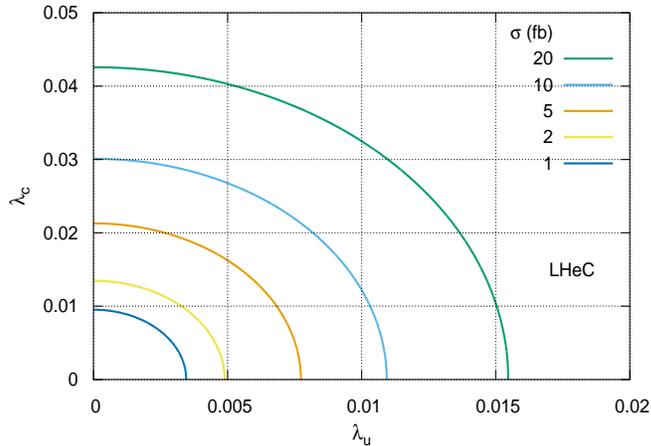}

\caption{Contour plot for plane of FCNC couplings $\lambda_{u}$ and $\lambda_{c}$
depending on signal cross sections at LHeC. \label{fig:fig2}}
\end{figure}

The signal cross sections for the process $e^{-}p\to(e^{-}t+e^{-}\bar{t})+X$
for different values of couplings $\lambda_{u}$ and $\lambda_{c}$
at the LHeC are given in Table \ref{tab:tab1}. In order to estimate
the sensitivity to FCNC coupling parameters in the range, we plot
contours corresponding to different values of signal cross sections
as shown in Fig. \ref{fig:fig2}. A realistic and detailed analysis
of the signal and background for one electron, two jets and one $b$-tagged
jet in the final state is given in the following section including
a fast detector simulation.

\begin{table}
\caption{The signal cross section values (in pb) for process $e^{-}p\to(e^{-}t+e^{-}\bar{t})+X$
at LHeC. \label{tab:tab1} }

\begin{tabular}{|c|c|c|c|}
\hline 
LHeC & $\lambda_{c}=10^{-2}$ & $\lambda_{c}=10^{-3}$ & $\lambda_{c}=0$\tabularnewline
\hline 
\hline 
$\lambda_{u}=10^{-2}$ & $9.468\times10^{-3}$ & $8.368\times10^{-3}$ & $8.357\times10^{-3}$\tabularnewline
\hline 
$\lambda_{u}=10^{-3}$ & $1.188\times10^{-3}$ & $9.460\times10^{-5}$ & $8.365\times10^{-5}$\tabularnewline
\hline 
$\lambda_{u}=0$ & $1.103\times10^{-3}$ & $1.104\times10^{-5}$ & $0$\tabularnewline
\hline 
\end{tabular}
\end{table}

\section{{\normalsize{}Analysis}}

In this section, we study the sensitivity to anomalous top FCNC couplings
through the signal and background process $e^{-}p\to e^{-}W^{\pm}q+X$
at LHeC. This process includes the off-shell single top quark production
and interfering background with the signal. We calculate the cross
section for this process to normalize the distributions from the signal
and background events. The cross section values depending on the FCNC
couplings are shown in Table \ref{tab:tab2} with the generator level
preselection cuts ($n_{e}\geq1$ and $n_{j}\geq$3). We generate signal
and background (interfering with signal) events by using MadGraph5\_aMC@NLO
\citep{Alwall11} with the effective Lagrangian implemented through
FeynRules \citep{Alloul14} model including the CKM matrix elements.
Parton showering and fast detector simulations are subsequently performed
with Pythia 6 \citep{Sjostrand06} and Delphes 3 \citep{deFavereau14},
respectively. For the analysis of signal and background events, following
the generator based pre-selection cuts, we also apply selection cuts
as shown in Table \ref{tab:tab3}. 

\begin{table}
\caption{The cross section (in pb) for the process $e^{-}p\to e^{-}W^{\pm}q+X$
depending on the FCNC couplings $\lambda_{u}$ and $\lambda_{c}$.
The values correspond to the case that only one of the couplings is
allowed to change at a time. \label{tab:tab2}}

\begin{tabular}{|c|c|c|c|c|c|}
\hline 
$\lambda_{u}$ or $\lambda_{c}$$\to$ & $0.05$ & $0.03$ & $0.02$ & $0.01$ & $0$\tabularnewline
\hline 
\hline 
$\lambda_{c}=0$ & 2.493 & 2.368 & 2.329 & 2.307 & 2.298\tabularnewline
\hline 
$\lambda_{u}=0$ & 2.324 & 2.308 & 2.303 & 2.299 & 2.298\tabularnewline
\hline 
$\lambda_{u}=\lambda_{c}$ & 2.519 & 2.378 & 2.333 & 2.307 & 2.298\tabularnewline
\hline 
\end{tabular}
\end{table}

\begin{table}
\caption{The preselection and a set of cuts for the analysis of signal and
background events at LHeC. \label{tab:tab3}}

\begin{tabular}{c||c||c||c||c||c||c||cl|ll}
\hline 
\multicolumn{8}{c}{Cut-0} &  &  & preselection : $n_{e}\geq1$ and $n_{jets}\geq3$\tabularnewline
\hline 
\multicolumn{8}{c}{Cut-1} &  &  & b-tag : one $b$-tagged jet ($j_{b}$)\tabularnewline
\hline 
\multicolumn{8}{c}{Cut-2} &  &  & transverse momentum : $p_{T}(j_{2},\,j_{3})>30$ GeV and $p_{T}(j_{b})>40$
GeV and $p_{T}(e)>20$ GeV\tabularnewline
\hline 
\multicolumn{8}{c}{Cut-3} &  &  & pseudo-rapidity : $-4<\eta(j_{b},\,j_{2},\,j_{3})<0$ and $|\eta(e)|<2.5$\tabularnewline
\hline 
\multicolumn{8}{c}{Cut-4} &  &  & $W$ boson mass : $50<M_{inv}^{rec}(j_{2},j_{3})<100$ GeV\tabularnewline
\hline 
\multicolumn{8}{c}{Cut-5} &  &  & top quark mass : $130<M_{inv}^{rec}(j_{b}+j_{2}+j_{3})<190$ GeV\tabularnewline
\hline 
\end{tabular}
\end{table}

\begin{figure}
\includegraphics[width=8cm,height=5cm]{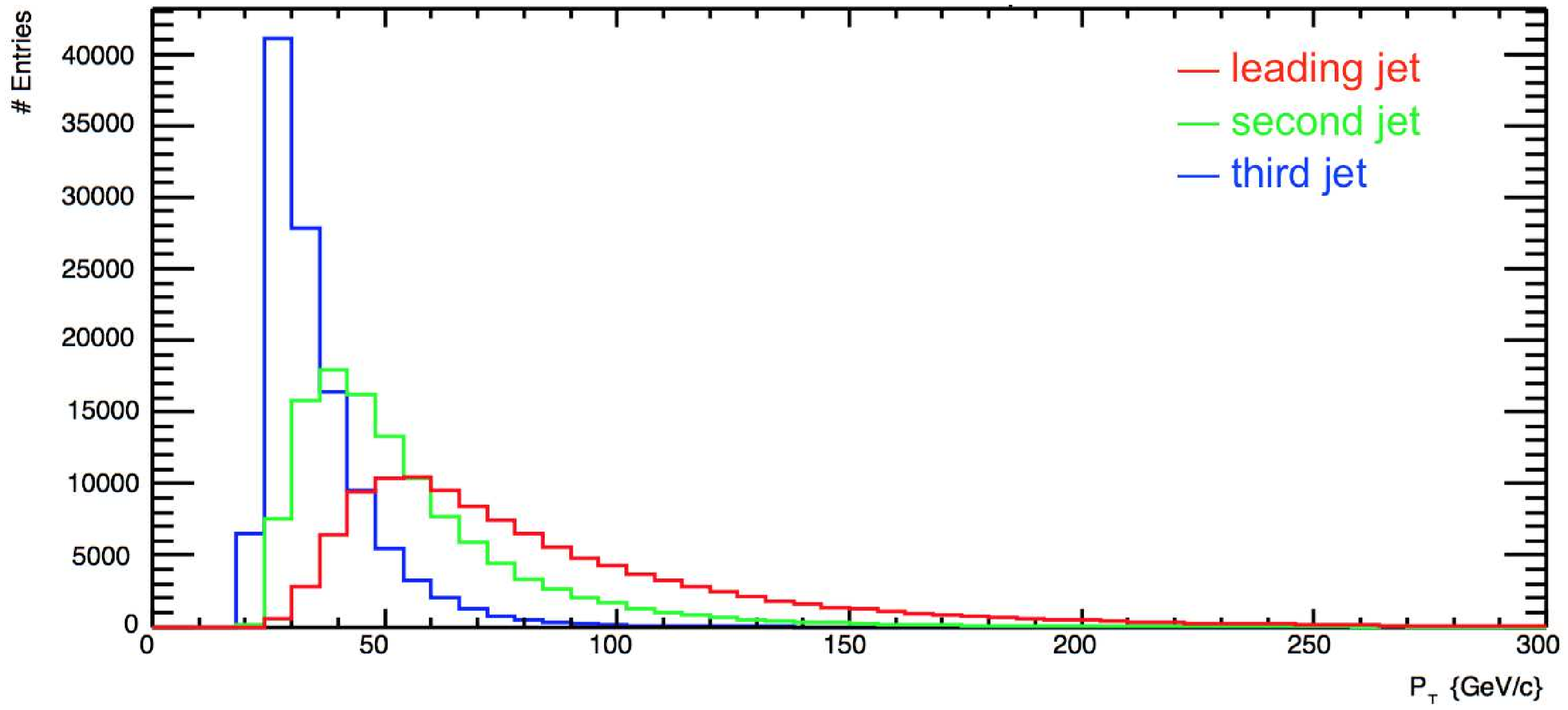}\includegraphics[width=8cm,height=5cm]{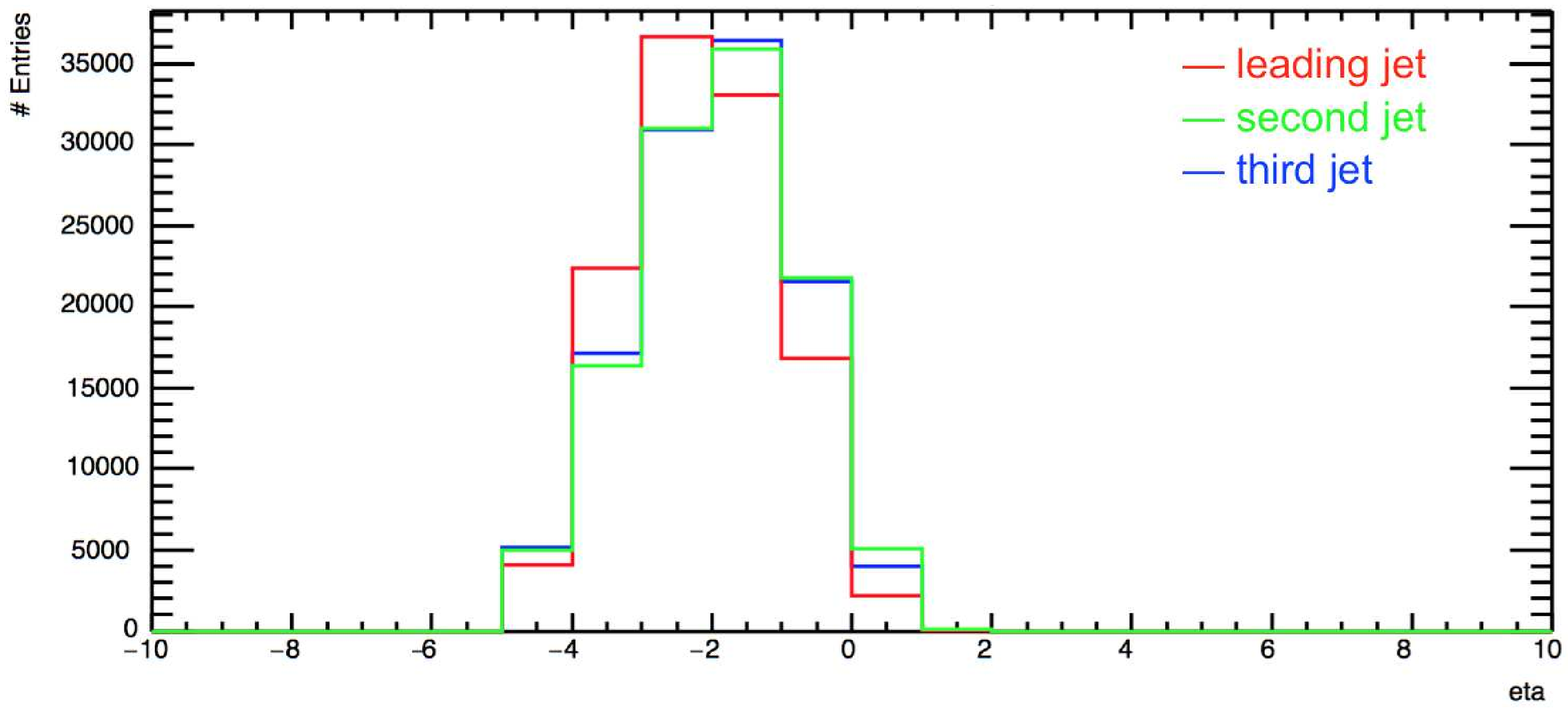}

\caption{Transverse momentum and pseudo-rapidity distributions of three jets
from the process $e^{-}p\to e^{-}W^{\pm}q+X$ which includes both
the interfering background and signal for $\lambda_{u}=\lambda_{c}=0.05$
at the LHeC. \label{fig:fig3}}
\end{figure}

\begin{figure}
\includegraphics[width=8cm,height=5cm]{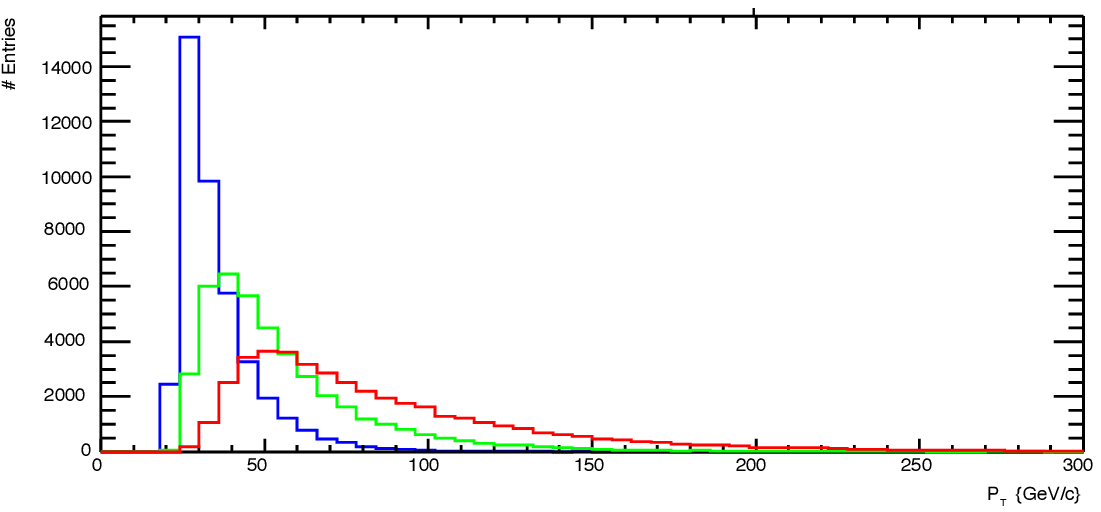} \includegraphics[width=8cm,height=5cm]{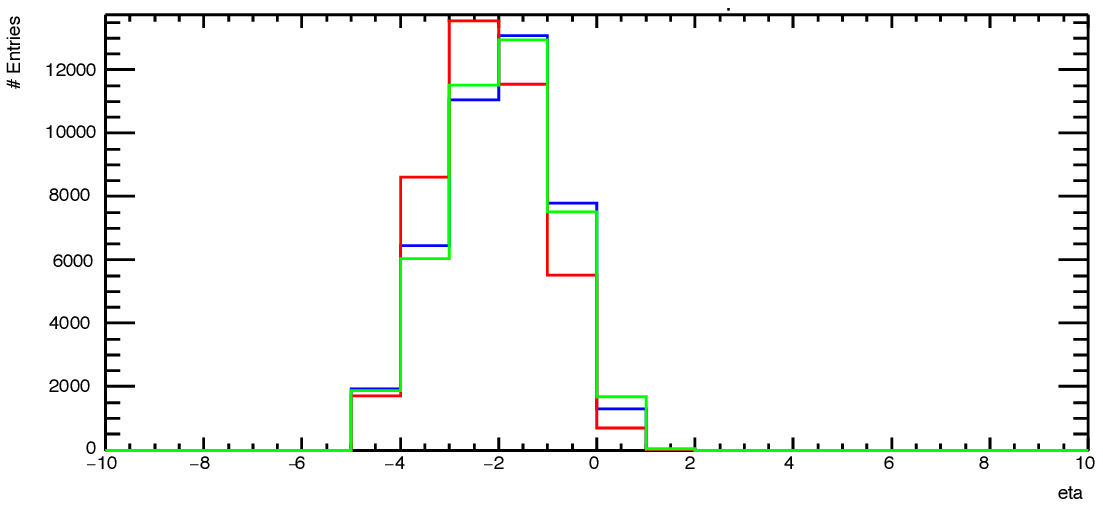}

\includegraphics[width=8cm,height=5cm]{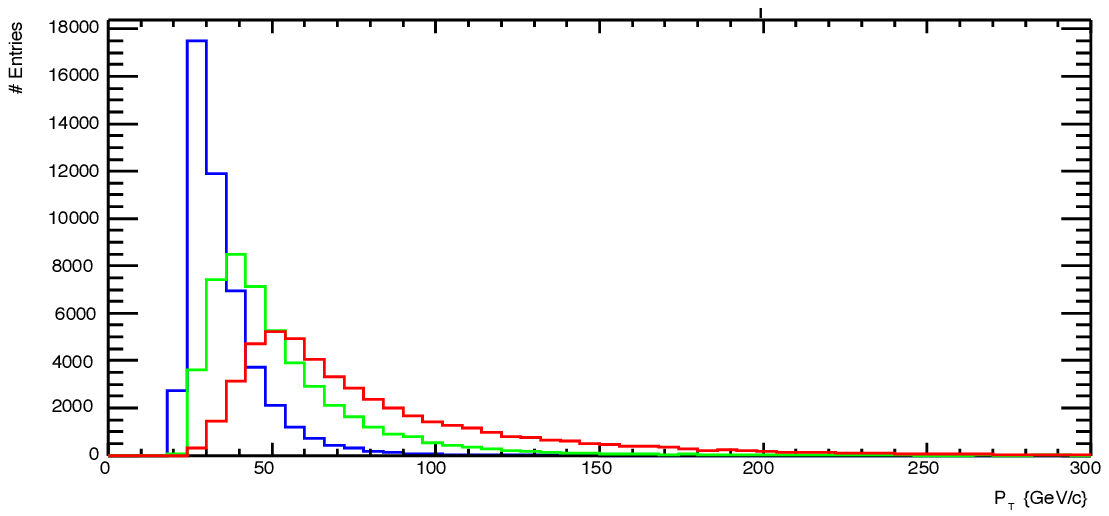}\includegraphics[width=8cm,height=5cm]{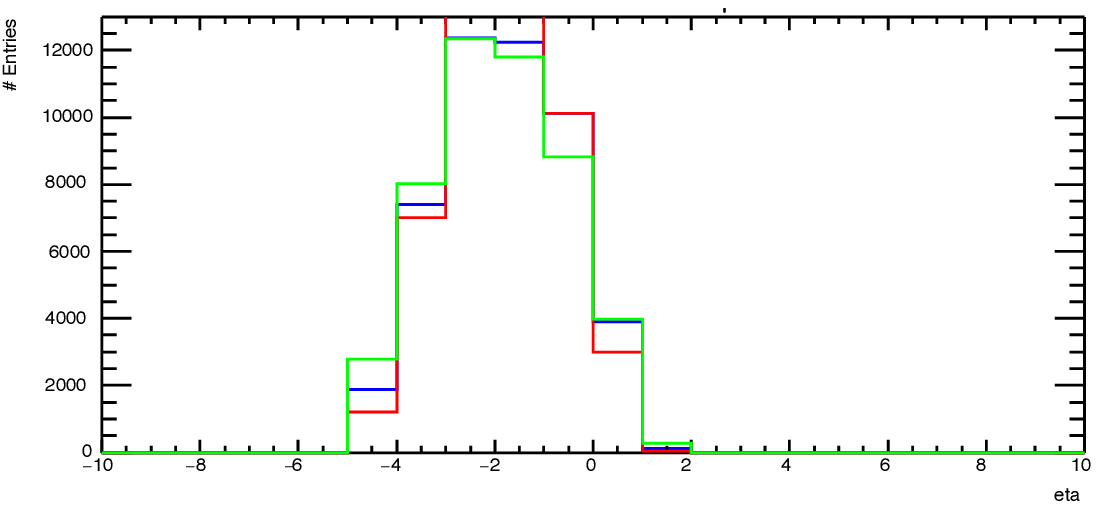}

\caption{Transverse momentum and pseudo-rapidity distributions of three jets
from the process $e^{-}p\to e^{-}W^{\pm}q+X$ for background (B1)
upper panels and the process $e^{-}p\to e^{-}Zq+X$ for background
(B2) lower panels at the LHeC. \label{fig:fig4}}

\end{figure}

\begin{figure}
\includegraphics[width=8cm,height=5cm]{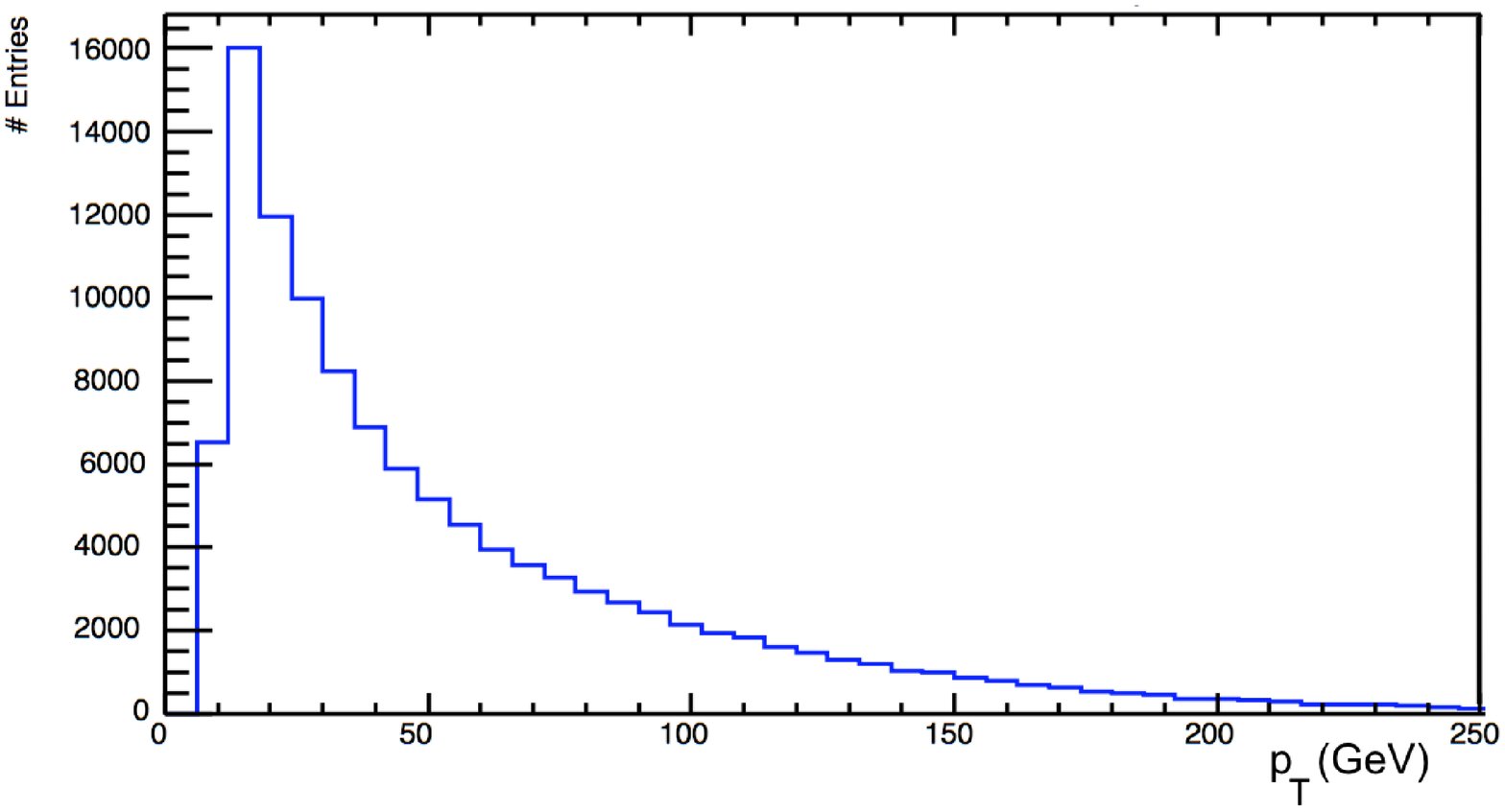}\includegraphics[width=8cm,height=5cm]{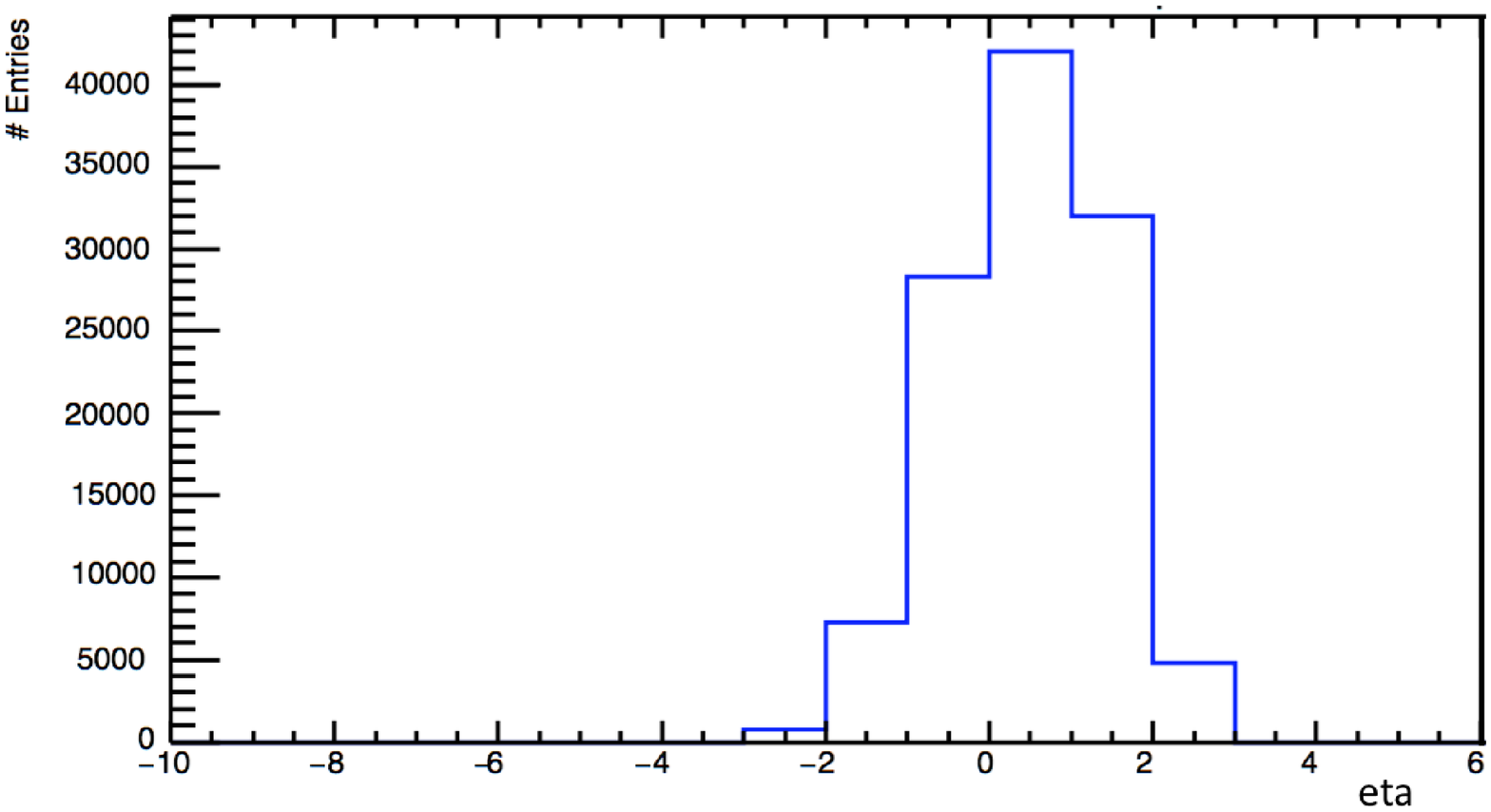}

\caption{Transverse momentum and pseudo-rapidity distribution of electron from
the process $e^{-}p\to e^{-}W^{\pm}q+X$ which includes both the background
and signal for $\lambda_{u}=\lambda_{c}=0.05$ at the LHeC. \label{fig:fig5}}
\end{figure}

\begin{figure}
\includegraphics[width=8cm,height=5cm]{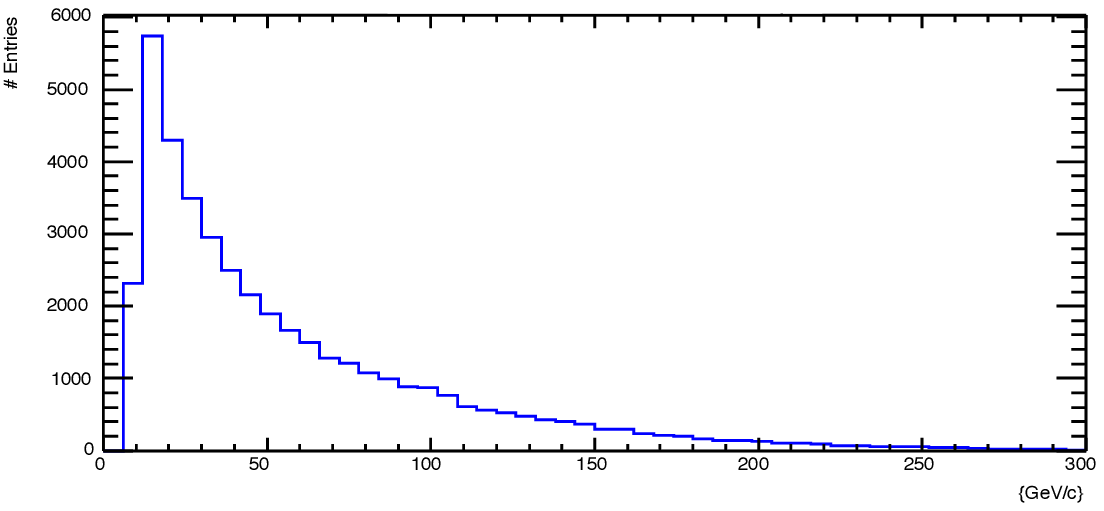}\includegraphics[width=8cm,height=5cm]{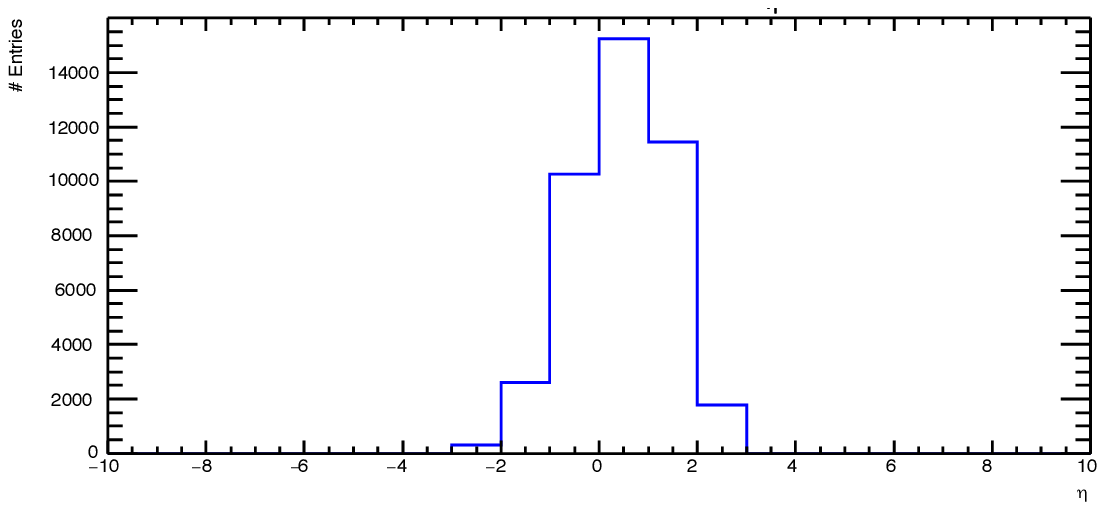}

\includegraphics[width=8cm,height=5cm]{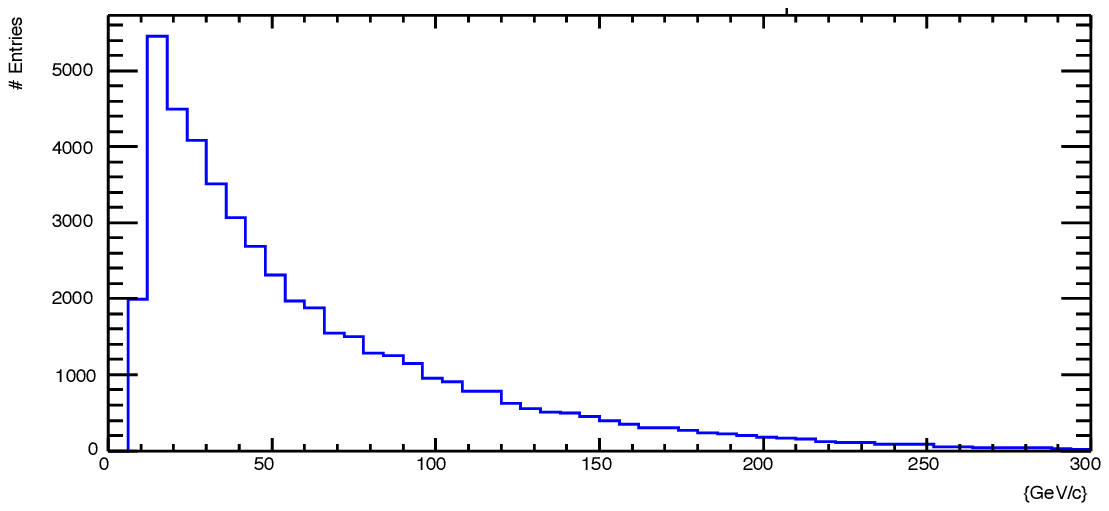}\includegraphics[width=8cm,height=5cm]{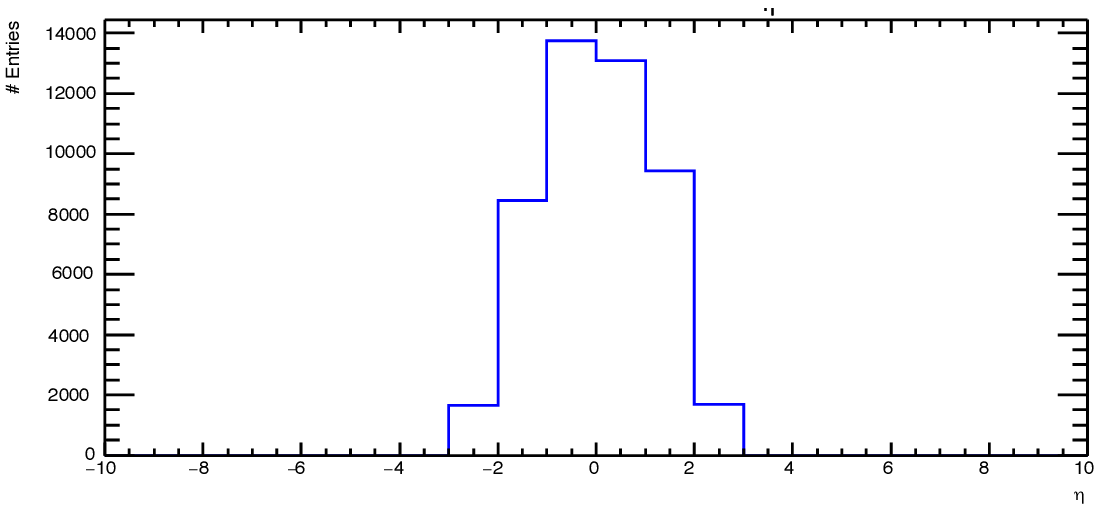}\caption{Transverse momentum and pseudo-rapidity distribution of electron from
the process $e^{-}p\to e^{-}W^{\pm}q+X$ for background (B1) upper
panels and the process $e^{-}p\to e^{-}Zq+X$ for background (B2)
lower panels at the LHeC. \label{fig:fig6}}
\end{figure}

Concerning our process $e^{-}p\to e^{-}W^{\pm}q+X$, with the hadronic
decay mode of the $W$ boson we will have at least three jets and
one electron in the final state. We take into account $eWq$ (B1)
and $eZq$ (B2) backgrounds which have the same final state after
W and Z decays hadronically, as that of the signal process. Here,
multi-jet QCD backgrounds are not included in the analysis. The transverse
momentum and pseudo-rapidity distributions of the leading jet, second
and third jets are presented in Fig. \ref{fig:fig3} for the signal
and interfering background together, and in Fig. \ref{fig:fig4} only
the backgrounds B1 and B2. The distribution have similar shape for
signal and consider backgrounds. In order to select signal events,
each event is required to have one electron, two jets and one $b$-jet
with the highest $p_{T}$. The number of selected objects in the signal
events can help to suppress the background events effectively. The
kinematical distributions ($p_{T}(e)$ and $\eta(e)$) for the electron
in the event are shown in Fig. \ref{fig:fig5} for the S+B1 and in
Fig. \ref{fig:fig6} for only B1 and B2. One of the specific signatures
of the signal is the presence of an electron with high transverse
momentum ($p_{T}(e)>20$ GeV) and central pseudo-rapidity ($|\eta(e)|<2.5$).
The leading jet with $p_{T}(j)>40$ GeV and other two jets with $p_{T}(j)>30$
GeV are considered in our analysis, as the cuts flow shown in Table
\ref{tab:tab3}. Since the LHeC has an energy asymmetry, the jets
from the process mainly peaked in the backward region according to
the $ep$ collisions, therefore the pseudo-rapidty range for jets
is taken to be $-4<\eta(j)<0$ for the analysis.

\begin{figure}
\includegraphics[scale=0.5]{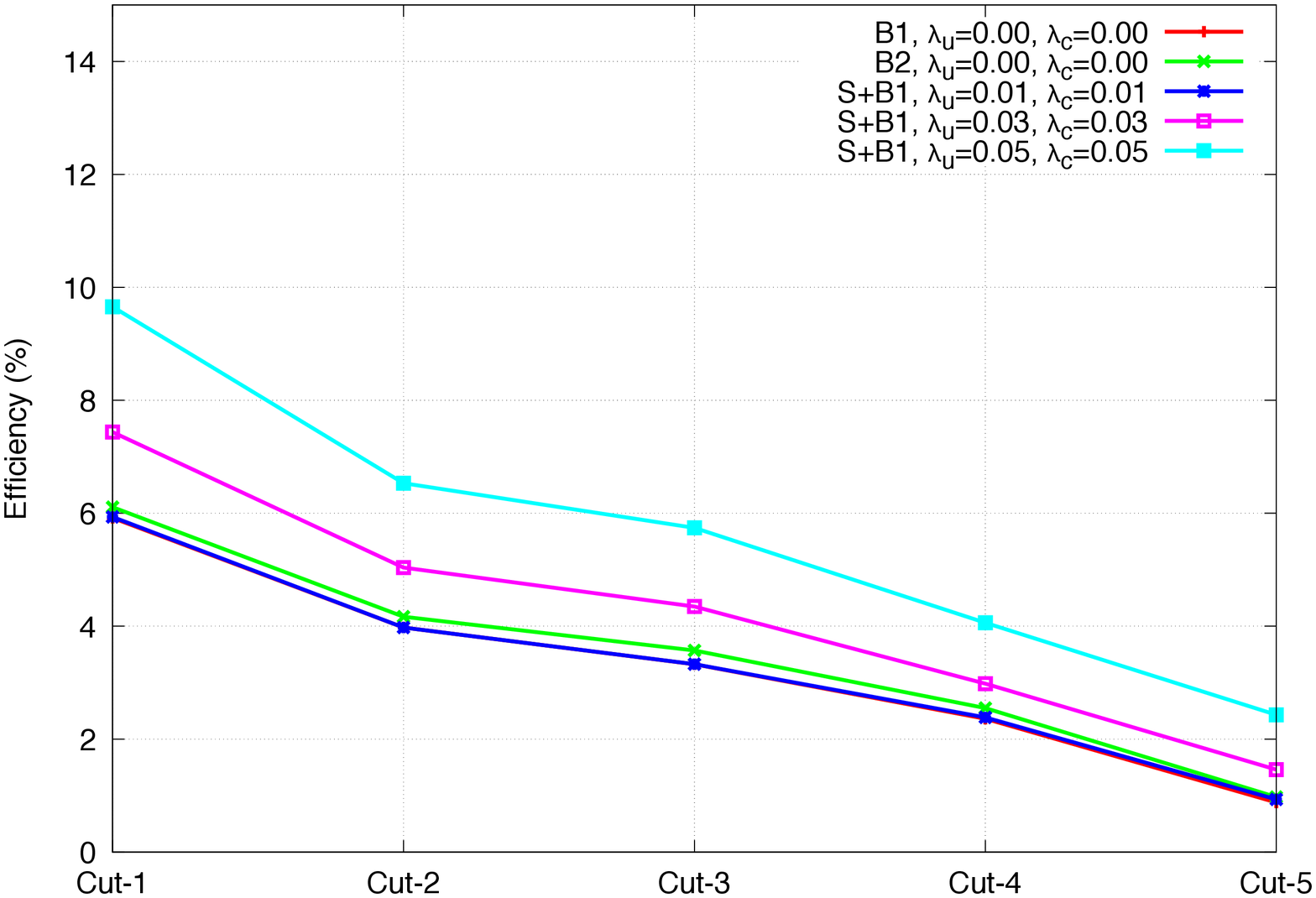}

\caption{Efficiency plot for the cuts applied at each step for the analysis
of signal (S)+background (B1) and background (B2) events. The cut
efficiencies are calculated with respect to the preselection cuts
for each coupling value. \label{fig:fig7}}
\end{figure}

In the fast simulation, the detector card includes $b$-tagging efficiency
to be $70\%$, with mistagging efficiencies for $c$-jets and light
jets of $10\%$ and $0.3\%$, respectively, in the considered ($p_{T},$
$\eta$) range of the jets. These are consistent with the typical
values used by the LHC experiments \citep{ATLAS2015}. The top quark
mass is reconstructed from three jets (with one $b$-tagged jet) after
selection cuts. At each cut step we calculate the cut efficiencies
as shown in Fig. \ref{fig:fig7} for different FCNC coupling parameter.
This figure shows that the cut efficiency changes from $6\%$ to $1\%$
for interfering background $eWq$ (B1) and $eZq$ (B2) with cut steps
from Cut-1 to Cut-5 when the FCNC coupling $(\lambda)$ is set to
zero. For the signal the cut efficiency depends on the value of the
couplings for example, it decreases from 10\% to 2.5\% for the couplings
$\lambda_{u}=\lambda_{c}=0.05$. 

\begin{table}
\caption{The number of events for background B1 (B2) and signal with FCNC couplings
$\lambda_{u}$ and $\lambda_{c}$ at LHeC with $L_{int}=100$ fb$^{-1}$.
\label{tab:tab4}}

\begin{tabular}{|c|c|c|c|c|}
\hline 
 & $\lambda_{c}=0$ & $\lambda_{c}=0.01$ & $\lambda_{c}=0.03$ & $\lambda_{c}=0.05$\tabularnewline
\hline 
\hline 
$\lambda_{u}=0$ & 584 (149) & 592  & 609 & 640\tabularnewline
\hline 
$\lambda_{u}=0.01$ & 617  & 621 & 692 & 763\tabularnewline
\hline 
$\lambda_{u}=0.03$ & 943 & 969 & 1003 & 1209\tabularnewline
\hline 
$\lambda_{u}=0.05$ & 1502 & 1744 & 1758 & 1792\tabularnewline
\hline 
\end{tabular}
\end{table}

The number of events for signal and backgrounds after Cut-5 at LHeC
with an integrated luminosity of $L_{int}=100$ fb$^{-1}$ are given
in Table \ref{tab:tab4}. The respective number of events after each
cut can be obtained from Table \ref{tab:tab4} with a relative factor. 

\begin{figure}
\includegraphics[scale=0.5]{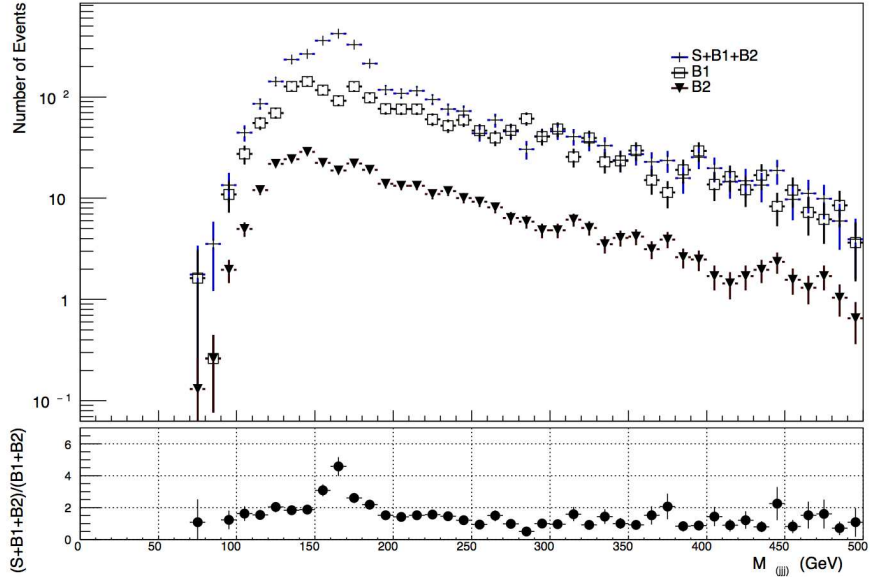}

\caption{Invariant mass distributions of three jets (one of the jets is required
as $b$-jet) for the signal+background (S+B1+B2), and backgrounds
(B1, B2). The ratio plot presents the signal (for equal coupling scenario
$\lambda=0.05$) strength which peaks at the top mass. \label{fig:fig8}}
\end{figure}

We plot the invariant mass of top quark (reconstructed from three
jets, one is $b$-tagged) as shown in Fig. \ref{fig:fig8}. The Cut-4
for $W$ boson mass window reduce $e^{-}Zq$ background (B2) effectively,
while keeps the signal less affected. The invariant mass ($M_{jjj}$)
distribution after Cut-4 and ratio (S+B1+B2)/(B1+B2) plot for equal
coupling scenario ($\lambda_{u}=\lambda_{c}=0.05$) are presented
in Fig \ref{fig:fig8}, where the main contribution comes from the
B1 background.

\begin{figure}
\includegraphics[scale=0.5]{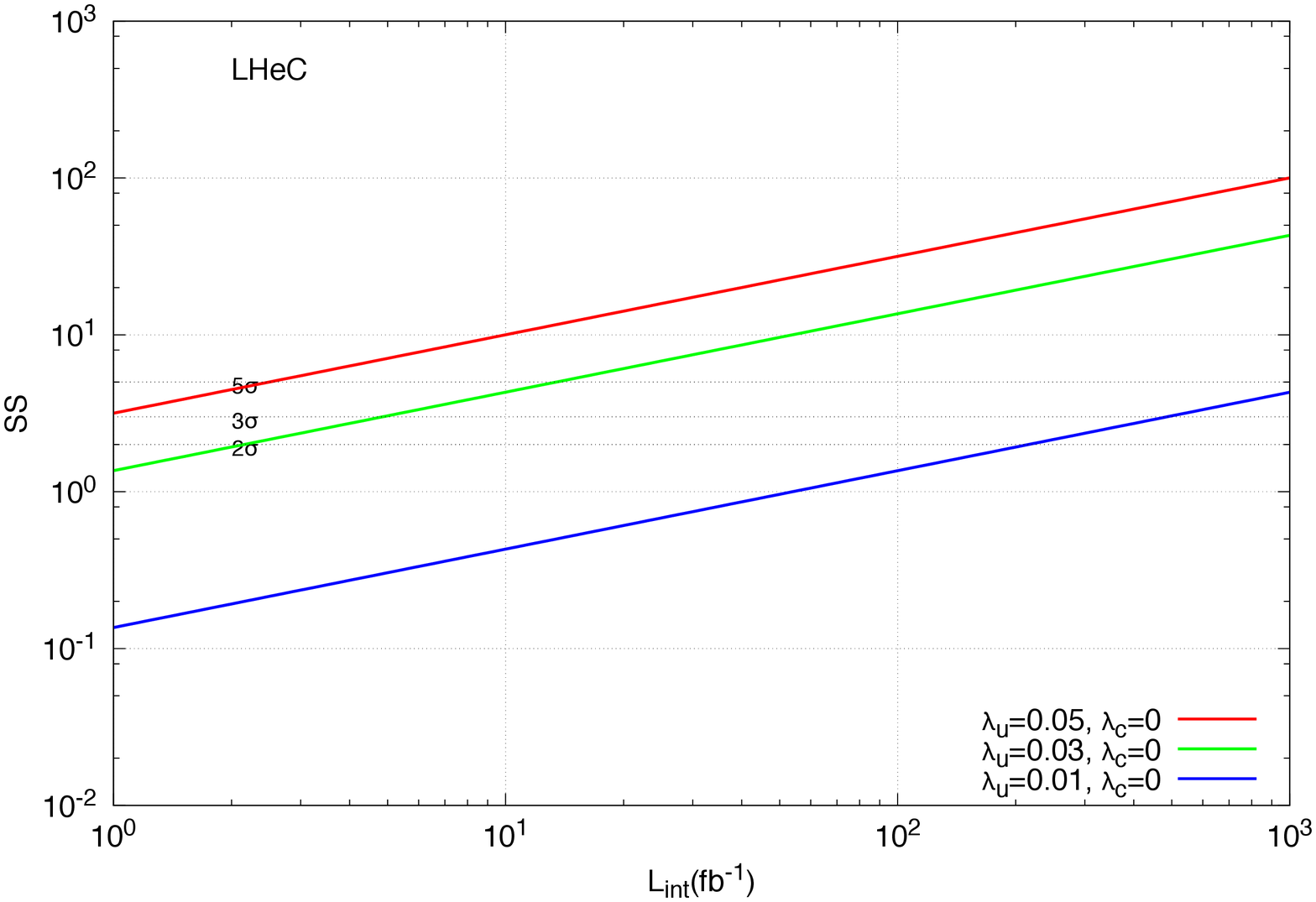}

\caption{Estimated statistical significance (SS) reach of flavor changing neutral
current $tu\gamma$ coupling ($\lambda_{u}$) depending on the integrated
luminosity ranging from 1 fb$^{-1}$ to 1 ab$^{-1}$ at the LHeC.
It includes the contribution from the main background (B1) on the
predicted results. The signal significance corresponding to $2\sigma$
, $3\sigma$ and $5\sigma$ lines are also shown. \label{fig:fig9}}
\end{figure}

\begin{figure}
\includegraphics[scale=0.5]{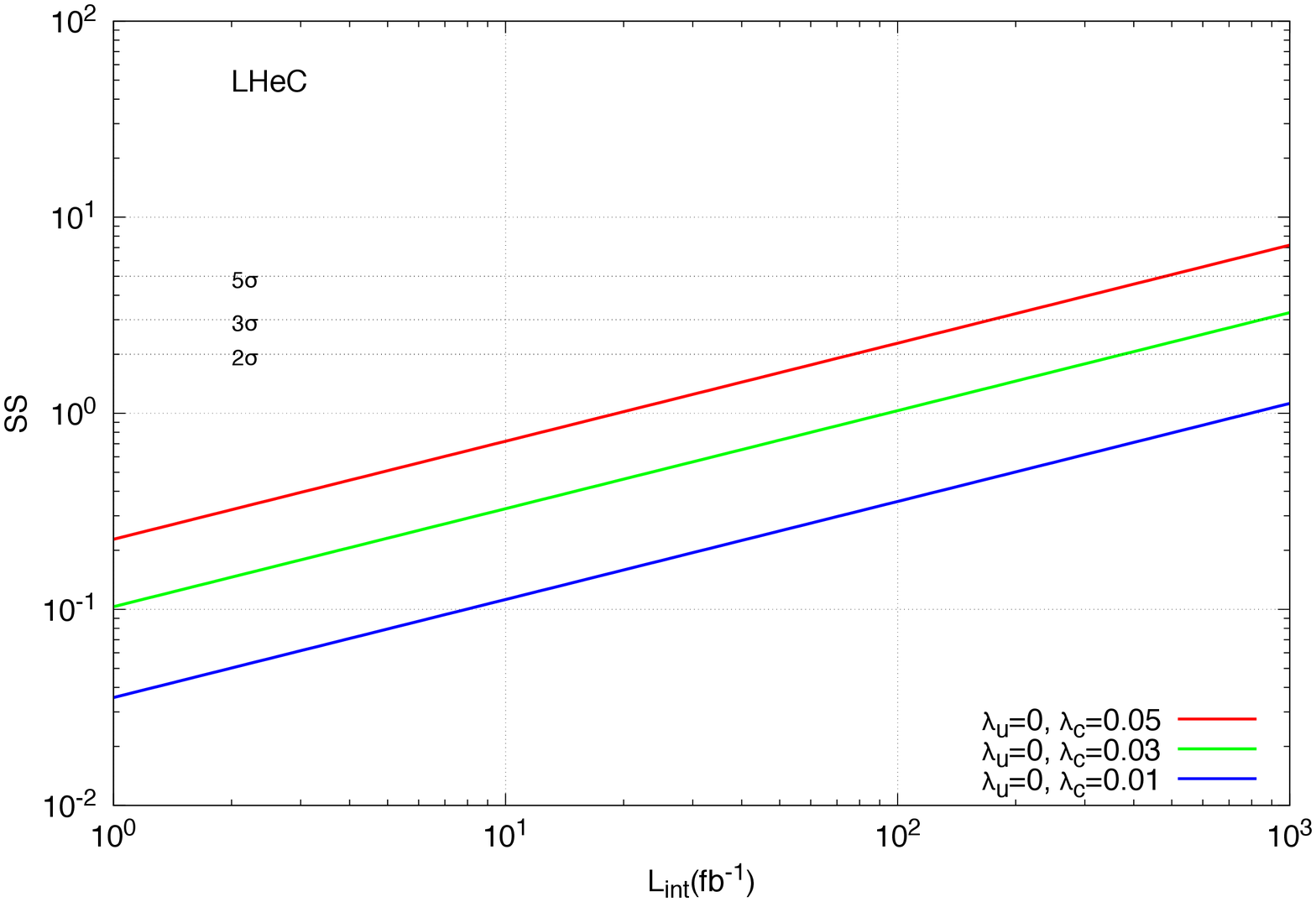}

\caption{Estimated statistical significance (SS) reach of flavor changing neutral
current $tc\gamma$ coupling ($\lambda_{c}$) depending on the integrated
luminosity ranging from 1 fb$^{-1}$ to 1 ab$^{-1}$ at the LHeC.
It includes the contribution from the main background (B1) on the
predicted results. The signal significance corresponding to $2\sigma$,
$3\sigma$ and $5\sigma$ lines are also shown. \label{fig:fig10}}
\end{figure}

\begin{figure}
\includegraphics[scale=0.8]{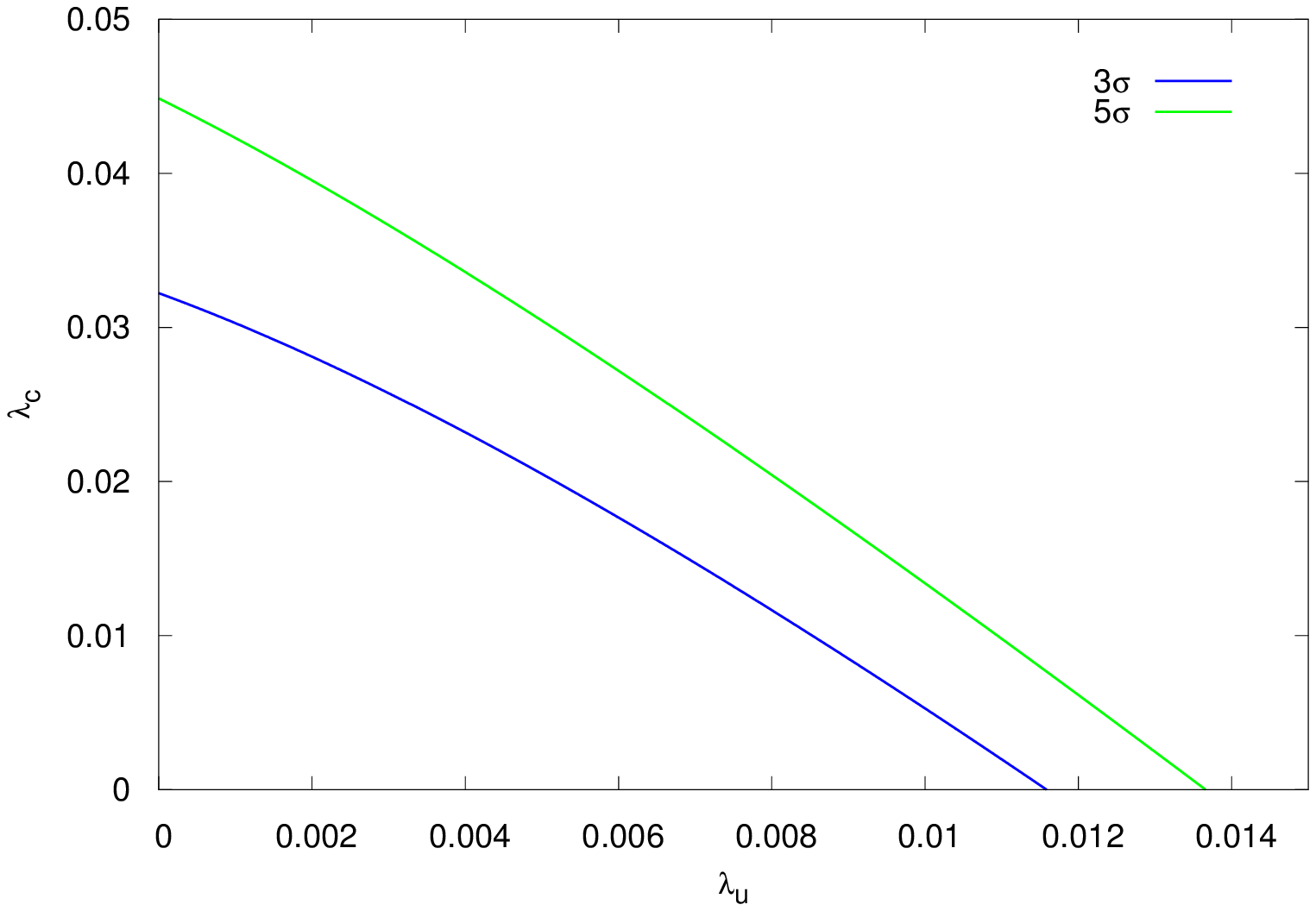}

\caption{The contour plot for the couplings $\lambda_{u}$ and $\lambda_{c}$
at LHeC for an integrated luminosity of 500 fb$^{-1}$. \label{fig:fig11}}
\end{figure}

The statistical significance ($SS$) are calculated at the final stage
of the cuts using the signal ($S$) and total background ($B$) events

\begin{equation}
SS=\sqrt{2[(S+B)\ln(1+\frac{S}{B})-S]}\label{eq:2}
\end{equation}
For different coupling scenario ($\lambda_{u}\neq\lambda_{c}$), we
calculate the significance and search for the sensitivity to the FCNC
couplings $\lambda_{u}$ and $\lambda_{c}$. The $SS$ reach for FCNC
$tq\gamma$ couplings are presented in Fig. \ref{fig:fig9} and Fig.
\ref{fig:fig10} depending on the integrated luminosity ranging from
$1$ fb$^{-1}$ to $1$ ab$^{-1}$ at the LHeC. It includes the contribution
from the main background on the predicted results. The signal significance
corresponding to $2\sigma$ , $3\sigma$ and $5\sigma$ lines are
also shown. In Fig. \ref{fig:fig11} we also present a plot showing
the $3\sigma$ and $5\sigma$ contours in the $\lambda_{u}-\lambda_{c}$
plane at LHeC for an integrated luminosity of 500 fb$^{-1}$. In this
study, we have calculated the cross section with the theoretical uncertainties
for the PDF scale variation which ranges $\pm2.5\:\%$. The LHeC collider
would be clear environment and the systematics would be less than
that of the LHC. For the sake of clarity, only uncertainties from
statistics are taken into account, while the uncertainties from systematics
are ignored in the present discussions.

In the literature, there are alternative use of effective coupling
constants appearing in the effective Lagrangian. It is also useful
to express our results in terms of branching ratios which can be comparable
with the results of other studies. In order to translate the bounds,
we define the branching ratio as 

\begin{equation}
BR(t\to q\gamma)=\frac{\Gamma(t\to q\gamma)}{\Gamma(t\to q'W^{+})+\Gamma(t\to u\gamma)+\Gamma(t\to c\gamma)}\label{eq:3}
\end{equation}
Since the $V_{tb}$ element of CKM matrix is much larger than $V_{ts}$
and $V_{td}$, the main contribution to $\Gamma(t\to q'W^{+})$ comes
from the decay $t\to bW$ for a value of about $\Gamma(t\to bW)=1.41$
GeV \citep{Patriagnani}. We calculate the partial widths for the
FCNC decay channels $t\to q\gamma$ as $\Gamma(t\to q\gamma)=(1/8)\alpha_{e}\lambda_{q}^{2}m_{t}$.

In this study, we find $3\sigma$ significance results to reach an
upper limits $\lambda_{u}=0.012$ and $\lambda_{c}=0.032$ with an
integrated luminosity of $500$ fb$^{-1}$ at LHeC. These limits on
the couplings can also be translated to the upper bounds on branching
ratios BR$(t\to u\gamma)<1.62\times10^{-5}$ and BR$(t\to c\gamma)<1.15\times10^{-4}$,
respectively. Previous experimental constraints on the branching fractions
of $t\rightarrow q\gamma$ (with $q$ representing an up or charm
quark) from the $ep$ collider HERA are $0.29\%$ from ZEUS \citep{ZEUS},
and $0.64\%$ from H1 \citep{H1} at the $95\%$ confidence level
(CL). At LHeC energy region the $c$-quark contribution to the process
becomes comparable with the $u$-quark contribution, hence the sensitivity
to $\lambda_{c}$ will be enhanced comparing to HERA results. We can
also compare our results with the recent results from the CMS experiment
with an integrated luminosity of $19.8$ fb$^{-1}$ which place an
upper bound on top quark FCNC branching ratios BR$(t\to u\gamma)<1.61\times10^{-4}$
and BR$(t\to c\gamma)<1.82\times10^{-3}$ at $95\%$ confidence level
\citep{CMS15}. Our results show that an order of magnitude improvement
can be achieved for the top quark FCNC branchings. 

\section{{\normalsize{}Conclusion}}

The LHeC projected as an ep collider, especially due to its respective
clean environment when compared to high energy hadron colliders and
the electron beam polarization possibility, will provide important
results for some certain processes. In this study, we focused on the
physics potential of the LHeC promoting its complementary and competitive
role. We have analyzed the process $e^{-}p\to e^{-}W^{\pm}q+X$ with
the signature including one isolated electron and one $b$-jet together
with two jets in the final state. The signal for this process includes
the top quark flavor changing neutral current couplings ($tq\gamma$)
through photon exchanges in electron-proton collisions. We set upper
limits on the top quark FCNC couplings from the analysis of signal
and background including detector effects through the fast simulation.
The expected limits on $tq\gamma$ couplings at HL-LHC have already
reported in Ref. \citep{ATLAS2013}, the branching ratios for $t\rightarrow q\gamma$
are $8\times10^{-5}$ and $2.5\times10^{-5}$ for $L_{int}=$ 300
fb$^{-1}$ and 3000 fb$^{-1}$ , respectively. The LHeC with the high
luminosity of 1 ab$^{-1}$ has the potential in probing the top FCNC
couplings ($\lambda_{u}$, $\lambda_{c}$), which can be comparable
or even better when compared to the bounds from the HL-LHC.

\section*{Conflicts of Interest}

The authors declare that there is no conflict of interest regarding
the publication of this paper.
\begin{acknowledgments}
We acknowledge illuminating discussions within the LHeC Higgs and
Top group. I. Turk Cakir would like to thank CERN for the hospitality,
where a part of this work has been completed.  
\end{acknowledgments}


\begin{thebibliography}{99}
\bibitem[1]{GIM70} S.L. Glashow, J. Iliopoulos, and L. Maiani, Phys.
Rev. D \textbf{2}, 1285 (1970). 

\bibitem[2]{Saavedra04} J.A. Aguilar-Saavedra, Acta Phys. Polon.
B \textbf{35}, 2695 (2004).

\bibitem[3]{Couture97} G. Couture, M. Frank, H. Konig, Phys. Rev.
D \textbf{56}, 4213 (1997).

\bibitem[4]{Lu03} G.R. Lu, F.R. Yin, X.L. Wang, L.D. Wan, Phys. Rev.
D \textbf{68}, 015002 (2003).

\bibitem[5]{CDF98} CDF Collaboration, Phys. Rev. Lett. \textbf{80},
2525 (1998).

\bibitem[6]{CMS15} CMS Collaboration, Report No: CMS-TOP-14-003,
CERN-PH-EP-2015-287 (2014). JHEP 1604 (2016) 035.

\bibitem[7]{Jeneret2008} J. de Favereau de Jeneret, S. Ovyn, Nucl.
Phys. Proc. Suppl. \textbf{179-180}, 277 (2008).

\bibitem[8]{Sun2014} H. Sun, Nucl. Phys. B \textbf{886}, 691 (2014),
arXiv:1402.1817 {[}hep-ph{]}.

\bibitem{Guo2016} Y.C. Guo, C.X. Yue, S. Yang, Eur. Phys. J. C\textbf{
76, }596 (2016).

\bibitem{Reza2015} R. Goldouzian, Phys. Rev. D \textbf{91,} 014022
(2015).

\bibitem{Denizli2016} H. Denizli, A. Senol, A. Yilmaz, I. Turk Cakir,
H. Karedeniz, O Cakir, Phys.Rev. D\textbf{96} (2017). 

\bibitem{LHeC} LHeC Study Group, J L Abelleira Fernandez \emph{et
al}, J. Phys. G: Nucl. Part. Phys. \textbf{39}, 075001 (2012).

\bibitem{Klein2016} M. Klein, Annalen Phys. \textbf{528}, 138 (2016).

\bibitem{J.L.Abelleira} J. L. Abelleira Fernandez, J. Phys. G: Nucl.
Part. Phys. \textbf{39} 075001, (2012).

\bibitem{Mukesh} M. Kumar, X. Ruan, and R. Islam, A.S. Cornell, M.
Klein, U. Klein, B. Mellado, Phy. Lett. B. \textbf{764}, 247-253 (2017).

\bibitem{Yuan2011} X. Yuan, Y. Hao and Y. D. Yang, Phys. Rev. D \textbf{83},
013004 (2011).

\bibitem{Li2011} X. Q. Li, Y. D. Yang, and X. B. Yuan, J. High Energy
Phys. \textbf{08}, 075 (2011).

\bibitem{Yang2014} Y. D. Yang and X. B. Yuan, Chin. Sci. Bull. \textbf{59},
3760 (2014).

\bibitem{Alwall11} J. Alwall, M. Herquet, F. Maltoni, O. Mattelaer,
T. Stelzer, JHEP \textbf{1106}, 128 (2011), arXiv:1106.0522 {[}hep-ph{]}.

\bibitem{NNPDF2013} R. D. Ball et al., NNPDF Collaboration, Nucl.
Phys. B \textbf{867}, Issue 2, 244\textendash 289 (2013).

\bibitem{Alloul14} A. Alloul, N. D. Christensen, C. Degrande, C.
Duhr, B. Fuks, Comput. Phys. Commun. \textbf{185}, 2250-2300 (2014),
arXiv:1310.1921 {[}hep-ph{]}.

\bibitem{Sjostrand06} T. Sjostrand, S. Mrenna, P.Z. Skands, JHEP
\textbf{05}, 026 (2006).

\bibitem{deFavereau14} J. de Favereau, C. Delaere, P. Demin, A. Giammanco,
V. Lemaitre, A. Mertens, M. Selvaggi, JHEP \textbf{1402}, 057 (2014).

\bibitem{ATLAS2015} ATLAS Collaboration, Expected performance of
the ATLAS b-tagging algorithms in Run-2, ATL-PHYS-PUB-2015-022. 

\bibitem{Patriagnani} C. Patriagnani et al., (Particle Data Group),
Chin. Phys. C 40, 100001 (2016).

\bibitem{ZEUS} ZEUS Collaboration, Phys. Lett. B \textbf{708} (2012)
27, arXiv:1111.3901. 

\bibitem{H1} H1 Collaboration, Phys. Lett. B \textbf{678} (2009)
450, arXiv:0904.3876.

\bibitem{ATLAS2013} ATLAS Collaboration, arXiv:1307.7292. 
\end{thebibliography}
\end{document}